# OPEN-COOPETITION IN THE PC AND MOBILE INDUSTRIES: THE WEBKIT CASE

Jose Teixeira - University of Turku - jose.teixeira@utu.fi

In an era of software crisis[1], the move of firms towards geographically-distributed software development teams is being challenged by collaboration issues. On this matter, the open-source phenomenon may shed some light, as successful cases on distributed collaboration in the open-source community have been recurrently reported (Bonaccorsi and Rossi, 2003; Raymond, 1999). While practitioners move with difficulties towards globally distributed software development, there is a lack of research addressing the collaboration dynamics of large-scale distributed software projects (Paasivaara and Lassenius, 2003; Sengupta et al., 2006). More particularly, even if there are empirical manifestations of collaboration among software-houses that market rival-software products within the same market, there is a clear lack of research addressing the development of information systems by coopetitive manners[2].

WebKit is an open-source project providing an engine that renders and interprets content from the World Wide Web. Its technology permeates our digital life since it can be found in the most recent computers, tablets and mobile devices sold by Apple, Google, Samsung, Nokia, RIM, HTC, and others. With more than 10 years of history, the WebKit project has brought together volunteers and firm-sponsored software developers that collaborate over the Internet by open and transparent manners while giving up the traditional intellectual property rights. Given the lack of understanding on the collaboration dynamics of large-scale distributed software projects in general, and the development of information systems by coopetitive manners in particular, we explored collaboration networks in the WebKit open-source project.

While addressing a previous call, from (Basole, 2009) for the advancement of methods and techniques to support the visualization of temporal aspects (e.g. pace, sequence) to represent change and evolution in ecosystems[3], we combined and virtual-ethnography (VA) with a Social Network Analysis (SNA) over publicly-available and naturally-occurring open-source data that allowed us to re-construct and visualize the evolution of the WebKit collaboration in a sequence of networks. We started by screening, by ethnographic manners, publicly available data such as company announcements, financial reports and specialized-press that allowed us to gain insights of the industrial context. After attaining a better understanding of the the competitive dynamics of the mobile-devices and PC industries, we started extracting

---

[1] A brief discussion on the software-crisis is provided by Fitzgerald, B. "Software Crisis 2.0." Computer 45.4 (2012): 89-91.

[2] The author knowledges valuable contributions from Information Systems research addressing joint-ventures and consortia settings. However, we did not find prior research within an explicit coopetition setting where rival-firms collaborate while marketing competing-products on the same market.

[3] Basole, R. employs the ecosystems term as a complex network of companies interacting with each other, directly and indirectly, to provide a broad array of products and services. Thus the ecosystem metaphor can also be applied in the WebKit project.

and analysing the social network of the WebKit community leveraging SNA (Scott, 2012; Wasserman and Faust, 1994), which is an emergent method widely established across disciplines of social sciences in general (Borgatti and Foster, 2003; Uzzi, 1996; Wasserman and Faust, 1994; Watts, 2004)

Sampling one of our network-visualizations, Figure 1, illustrates the latest phase of the WebKit project from the end of 3 February 2011 (i.e. Nokia and Microsoft's announcement of a strategic partnership) to 3 April 2013 (i.e. Google forks the WebKit core creating the Blink project). We can observe that contributors sponsored by Nokia and Intel are on opposite sides of the network, reflecting the lack of collaboration between those two firms in the WebKit project.

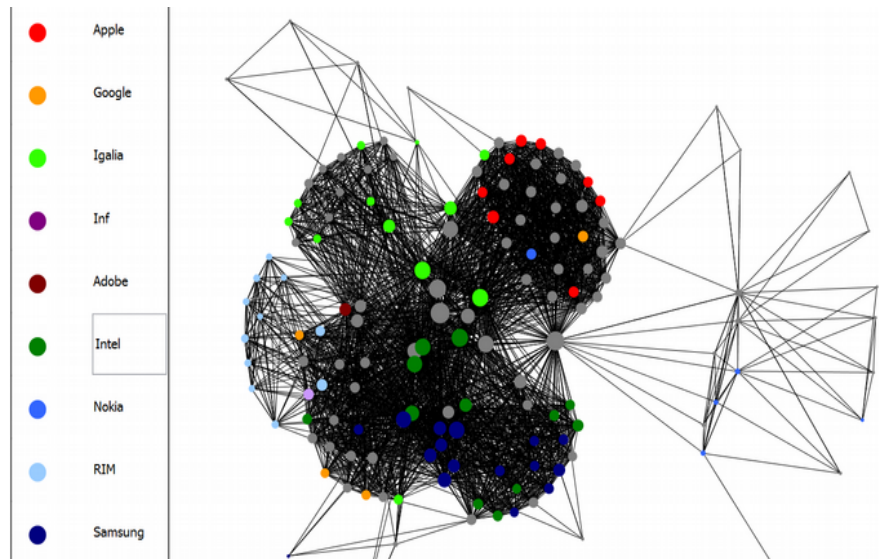

*Figure 1      Patent-wars, trademarks and forking (February 2011 to April 2013)*

More interestingly, by observing Apple and Samsung's roles, the orange and dark-blue nodes within Figure 1 network, we also attain interesting findings: Even if Samsung and Apple are involved in expensive patent wars in the courts (Bloomberg, 2013) and stopped collaborating on hardware components (Korea Times, 2013), their contributions remained strong and central within the WebKit open-source project. Even if the intense rivalry between Apple and Samsung is recognized officially by both companies and their law-attorney representatives; and well documented by the generalist and technological-specialized press; they software developers still collaborate intensively with each other in the WebKit open-source community.

Our case confirmed much of the established literature on coopetition. The need for external resources is the main driving force behind establishing long-term cooperative relationships to secure access to unique resources (Kock, 1991). Through cooperation, two companies can gain access to the other firm's unique resources or share the cost of developing new unique resources (Bengtsson and Kock, 2000). Within an open-source scenario, it is a networked community integrating firms and independent developers, open to contributions from everyone, that fulfils the need for external resources. Moreover, according to (Bengtsson and Kock, 2000), individuals within a firm can only act in accordance with one of the two logics of interaction at a time, either compete or collaborate. Hence that, either the two parts have to be divided between individuals within the company, or that one part needs to be controlled and regulated by an intermediate organization such as a collective association. Within an open-source scenario, it is the project community that plays the role of such intermediate organization. Developers must identify themselves with the project community for being able to collaborate with rivals in the same community.

Even if our initial observations complied with the classical coopetition management theories, more detailed observations revealed that coopetition theories can't be generalised to the open-source arena. Coopetition theories lack explanation power for the competitive and collaborative issues that are simultaneously present and interconnected in our WebKit case in particular, and the open-source community in general. First of all, most of classical coopetition theories derive from partnerships between two organizations while open-source coopetitive relationships tend to be more networked; exceptions, such as (Dagnino and Padula, 2002) take a inter-firm networked approach to coopetition, but based on in buy-sell relationships that do not make sense in the open-source community where many contribute in a voluntary basis. Second, must of the established literature on R&D coopetition address alliances that take the form of either joint-ventures or consortia, where access is only granted to a few selected partners; this contrasts with the open-source community where everyone is welcome to contribute to the project and everyone is allowed to copy, sell and distribute outcomes from the project. Finally, classical coopetition literature argues that coopetition activities take place far from the customer; competitors cooperate with activities far from the customer and compete in activities close to the customer (Bengtsson and Kock, 2000). However, in the open-source arena, coopetition can also happen very close to the customer in the terms of user-innovation (Von Hippel, 2009; Lakhani and Von Hippel, 2003)

Therefore we propose the development of a new theory on open-competition. A fork from the current theoretical body of knowledge on coopetition, that should explain the development of R&D projects by open-source manners in a high-networked environment. A theory that should explain collaboration between rival-firms by open-source manners, under public domain, where nobody needs to ask for permission to contribute or innovate; all within an open intellectual property regime where everyone is allowed to modify, copy, sell and distribute community-driven innovations.

***Keywords:*** *Collaboration, open-source, distributed software development, coopetition, open-coopetition, strategic alliances*